\documentclass[%
 aip,
 amsmath,amssymb,
preprint,%
 reprint,%
floatfix
]{revtex4-2}
\usepackage{amsfonts}  
\usepackage{amsmath}
\usepackage{natbib}
\usepackage{graphicx}
\usepackage{xcolor}
\usepackage{pifont}
\newcommand{\xmark}{\ding{55}}%

\usepackage{dcolumn}
\usepackage{bm}

\usepackage[T1]{fontenc}
\usepackage[utf8]{inputenc}
\usepackage{mathptmx}
\usepackage{etoolbox}
\usepackage{relsize}
\makeatletter
\def\@email#1#2{%
 \endgroup
 \patchcmd{\titleblock@produce}
  {\frontmatter@RRAPformat}
  {\frontmatter@RRAPformat{\produce@RRAP{*#1\href{mailto:#2}{#2}}}\frontmatter@RRAPformat}
  {}{}
}%
\makeatother

\usepackage{hyperref}
\hypersetup{
    colorlinks=true,
    linkcolor=blue,
    filecolor=magenta,      
    urlcolor=cyan,
    anchorcolor = black,
    pdftitle={Overleaf Example},
    pdfpagemode=FullScreen,
    }
\begin{document}

\preprint{AIP/123-QED}

\title{Recognizing and generating knotted molecular structures by machine learning}

\author{Zhiyu Zhang$^{\#}$}
\author{Yongjian Zhu$^{\#}$}
\author{\, Liang Dai{*}}

\affiliation{%
Department of Physics, City University of Hong Kong, Hong Kong SAR, China
}%
 \email{liangdai@cityu.edu.hk}
\date{\today}

\begin{abstract}
Knotted molecules occur naturally and are designed by scientists to gain special biological and material properties. Understanding and utilizing knotting require efficient methods to recognize and generate knotted structures, which are unsolved problems in mathematics and physics. Here, we solve these two problems using machine learning. First, our Transformer-based neural network (NN) can recognize the knot types of given chain conformations with an accuracy of $>99\%$. We can use a single NN model to recognize knots with different chain lengths, and our computational speed is about 4500 times faster than the most popular mathematical method for knot recognition: the Alexander polynomials. Second, we for the first time design a diffusion-based NN model to generate conformations for given knot types. The generated conformations satisfy not only the desired knot types, but also the correct physical distributions of the radii of gyration and knot sizes. The results have several implications. First, the Transformer is suitable for handling knotting tasks, probably because of its strength in processing sequence information, a key component in knotting. Second, our NN can replace mathematical methods of knot recognition for faster speed on many occasions. Third, our models can facilitate the design of knotted protein structures. Lastly, analyzing how NN recognizes knot types can provide insight into the principle behind knots, an unsolved problem in mathematics. We provide an online \href{http://144.214.24.236/}{website} for using our models.
\end{abstract}


\maketitle
\section{Introduction}

Knotting is a common phenomenon in everyday objects and chain-like molecules, including DNA \cite{Krasnow1,Rybenkov1,Rybenkov2,Arsuaga1,Plesa1} and proteins \cite{Taylor1,Mallam1,San1,Rubach1}. It has attracted a broad range of scientists because knotting can impact many systems. In biology, DNA knots are involved in many key biological processes \cite{Sogo1,Liu1,Marenduzzo1,Vologodskii1,Schvartzman1}, such as DNA replication, gene expression and cell division. DNA knots must be untied by topoisomerases in cells, otherwise, cells will die \cite{Berger1}. Hence, many cancer drugs kill cancer cells through inhibiting topoisomerase \cite{Nitiss1}. More than one thousand knotted protein structures have been discovered \cite{Jamroz1,Rubach1}. Why natural evolution chooses knotted proteins remains unclear. Some possible reasons are enhancing structural stability \cite{Sulkowska1,zhu2022computational} and bringing special catalytic functions through knotting \cite{Christian1}. Similarly, chemists found that a manmade knotted structure has special catalytic functions \cite{Marcos1}. 

In polymer physics, knotting can significantly slow down the relaxation of a compressed polymer \cite{Tang1} and the stretching kinetics of a polymer \cite{Renner1}. In addition, knotting can reduce the minimum breaking force of a polymer \cite{Zhang3}. From the practical viewpoint, knotted structures have been suggested for controlled drug release because untying knots can release the substances wrapped by knots \cite{Coluzza1}. In nanotechnology, DNA knots occur in nanopores \cite{Plesa1,Sharma1,Sharma2,Suma1} and nanochannels \cite{Reifenberger1,Mao1,Ma1}, which are designed for DNA analysis. Knotting may be utilized to achieve technical needs, such as slowing down nanopore translation for better data interpretation.

\begin{figure*}
\centering
\includegraphics[width=0.95\textwidth]{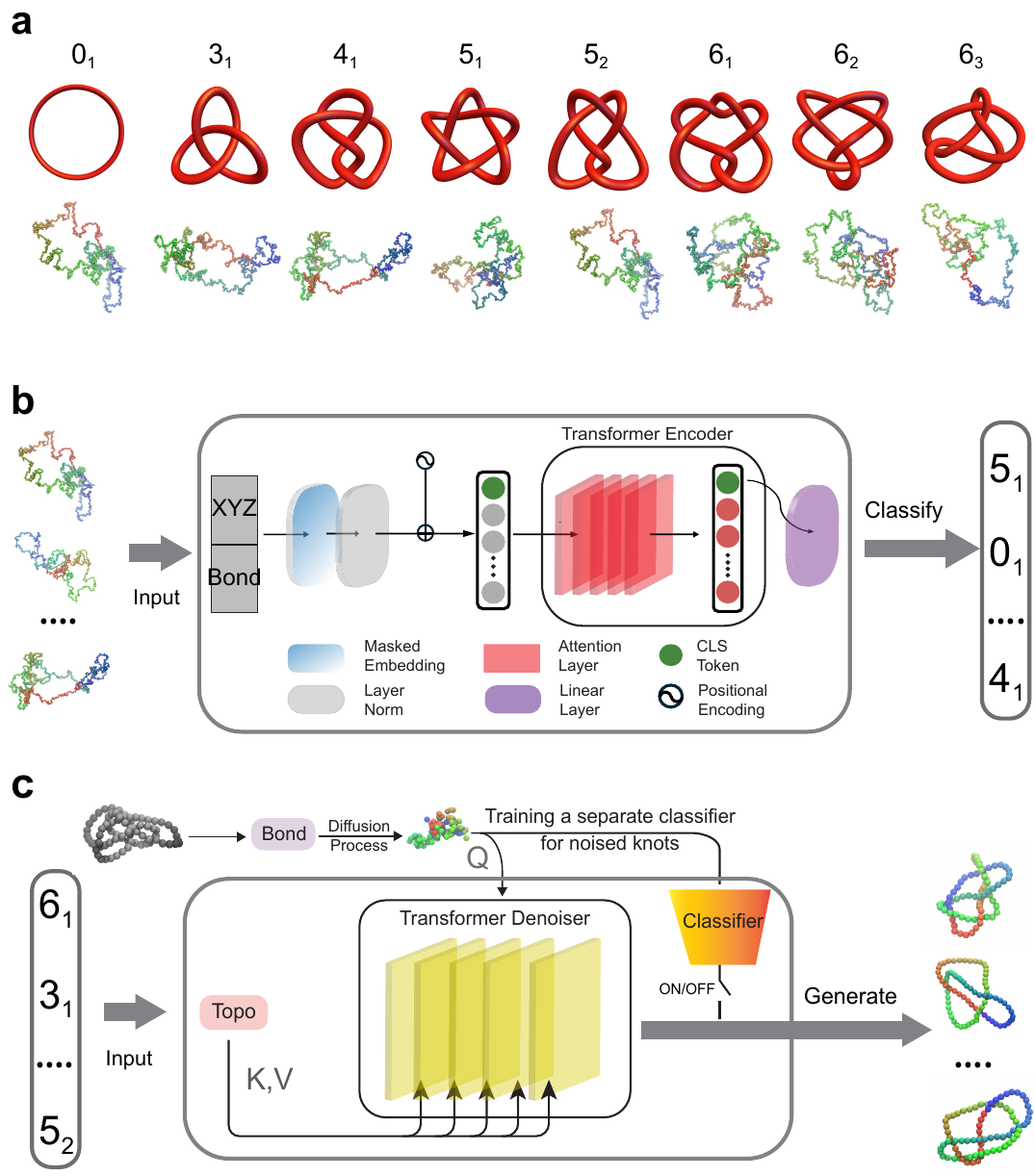}
\caption{\label{fig:Schematics}A schematic diagram of our models. (a) Long flexible polymer knots with their corresponding simplified sketch. First row from Left to Right: unknot ($0_1$), trefoil knot ($3_1$), figure-8 knot ($4_1$), and Cinquefoil knot ($5_1$), three-twist knot ($5_2$), Stevedore knot ($6_1$), $6_2$ knot, and $6_3$ knot. Due to thermal fluctuations, the unit bond vector of each segment does not correlate with their neighboring segment, resulting in a jiggling configuration. (b) Classification model architecture. Knotted flexible polymers and knotted proteins of various lengths are first embedded with padding. A class token (CLS token) is concatenated with the resulting embedding features for final classification. We use the vanilla Transformer encoder for cross-attention among polymer beads before a final linear layer. (c) Generation model architecture. Bond vectors of knotted semi-flexible polymers are subject to the noise scheduling for the Transformer blocks that learn to denoise. The input noised bond vector is cross-attended to the topological invariant inside the Transformer Denoiser. A separate classifier is trained to recognize the topology of noised knots and can be switched on and off during the generation process to guide synthesis by skewing the learned mean in conditional diffusion model.}
\end{figure*}

Two foundations in the research of physical knotting are recognizing and generating knotted conformations. Recognizing a knotted conformation means identifying the knot type of a given conformation. This is a difficult task, in principle, not fully solved in math. Fig.\,\ref{fig:Schematics}a shows a few conformations and their knot types. Recognizing their knot types by human eyes is impossible. 

Mathematicians have developed methods to recognize the knot types in the past century, including polynomial invariants (Alexander, Jones, and others) \cite{Alexander1,Jones1,Freyd1}, and homology invariants \cite{Ozsváth1,Khovanov1}. Knot invariants are defined based on the fact that for a closed rope, a quantity remains unchanged no matter how we distort the rope. This quantity is the invariant or knot type.  This invariant is often a polynomial, e.g. $t-1+t^{-1}$ for $3_1$ knot. A perfect knot-recognition method should accomplish two tasks: (i) for all conformations with the same knot type, the corresponding invariants (polynomials) are same; (ii) for conformations with different knot types, the corresponding polynomials are different. However, no method can accomplish the second tasks so far. For example, the method of the Alexander polynomials gives the same polynomial for the unknot and many other knot types. 

 Recent studies have applied machine learning to recognize knot types for given conformations. In 2020, we used the Long Short-Term Memory (LSTM) model \cite{Hochreiter1} and achieved an accuracy of 99\% for polymer conformations with chain lengths of $N=100$ and five knot types: $0_1$, $3_1$, $4_1$, $5_1$, and $5_2$ \cite{Vandans1}. The input feature is the bond vectors of polymer conformations. Later, Orlandini group \cite{Braghetto1} , Sulkowska group \cite{Sikora1}, Michieletto group \cite{Sleiman1}, and Lu group \cite{Wang1} performed similar studies using different neural networks (NNs), different input features, and different chain lengths. The maximum chain length in most of these studies is $N=200$. One practical application of these models is to hopefully replace the mathematical methods of knot recognition in molecular knot research, such as the popular Alexander polynomial method in recent studies \cite{Tubiana1,Orlandini1,Dai1}. NN models are often faster than mathematical methods in the knot recognition task. However, one limitation of all previous NN models is that they are trained for a fixed short chain length and can only be applied to that specific length due to RNN's inability to learn long-term dependency \cite{Bengio1}. This means that it is nearly impossible if one wants to use RNN for long polymer knot classification.
 
To address the above chain length-generalization problem, we revolutionize the current RNN architecture with the Transformer \cite{Vaswani1}. Originally developed for Natural Language Processing, the Transformer model has demonstrated exceptional performance in tasks such as recognition \cite{Dosovitskiy1} and text generation \cite{Radford1,Brown1}. The reasoning behind our choice is the long-range learning enabled by the attention mechanism, which is paramount in capturing the topology of knots on the long polymer chain. Also, its parallel processing capability allows for fast computation, beating the current algorithm in calculating the Alexander polynomial. Using Transformer as our base architecture, we train a model for a wide range of chain lengths from 100 to 1000, which covers most chain lengths in the recent simulation studies of DNA knots \cite{Mao1}, polymer knots \cite{Tubiana1,Orlandini1,Dai1}, and most lengths of natural protein knots \cite{Jamroz1,Rubach1}. In addition, we expand the number of knot types to eight: $0_1$, $3_1$, $4_1$, $5_1$, $5_2$, $6_1$, $6_2$, and $6_3$, which also covers the most knot types in recent studies. Furthermore, our model achieves a higher accuracy than the LSTM models in previous studies. 

In addition to improving knot recognition, we for the first time apply machine learning to generate knotted structures in this work. Previous studies have largely overlooked the generation of desired topology, a critical aspect for applications in AI protein design \cite{Trippe1,Ingraham1,Notin1} and molecular simulation\cite{Zhang2,Noe1,Arts1,Hsu1}. Generating polymers that adhere to physical and topological constraints is essential, particularly as knotted polymers are increasingly recognized for their biological functions and associations with diseases \cite{Virnau1,Kolesov1,San1,Lou1}. In this work, the generated conformations from our diffusion-based NN not only satisfy the knot type but also display the correct distributions of knot size and radii of gyration, which are crucial indicators in future AI-driven knotted protein design and molecular simulation.

\section{Results}
\subsection{Recognizing knot types by machine learning}


\begin{figure*}
\centering
\includegraphics[width=\textwidth]{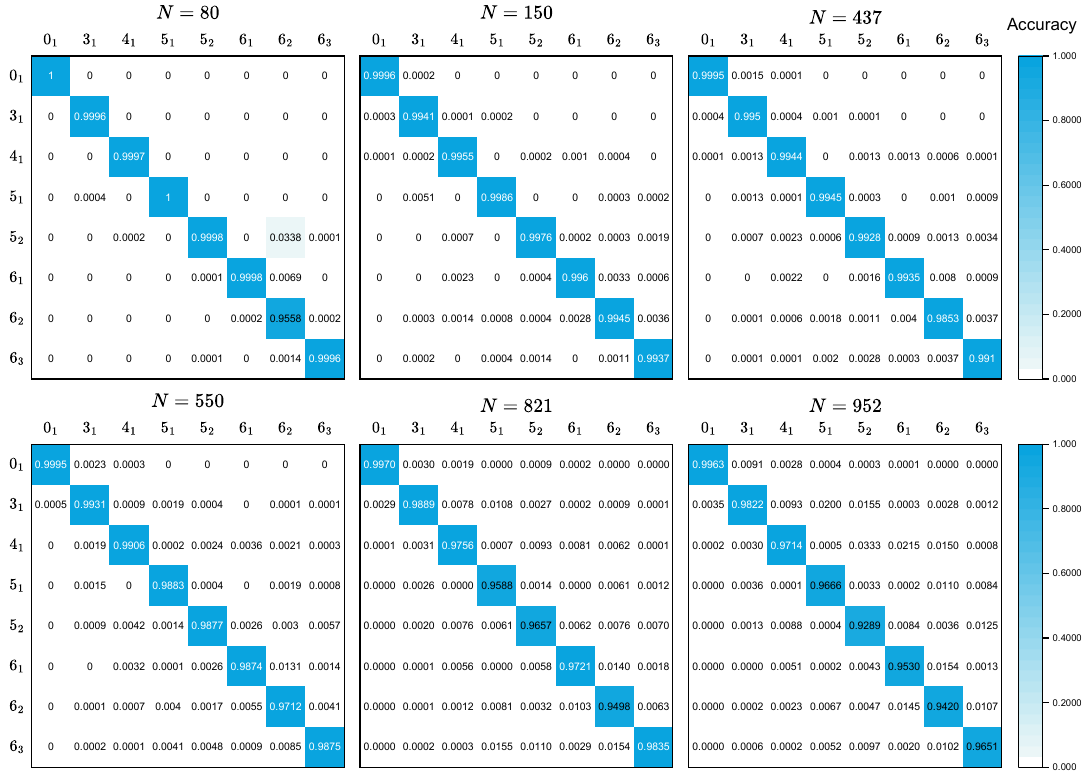}
\caption{\label{fig:bigmodel}The confusion matrices of our classification model. Six matrices correspond to six polymer lengths:  $N=80,\,150,\,437,\,550,\,821,\,952$, which are arbitrarily chosen and are not included in the training set. For all matrices, the horizontal coordinate represents the actual ground truth class, while the vertical coordinate represents the predicted.}
\end{figure*}

\begin{table*}
\caption{\label{tab:table1}A comparison between the existing models and our new model with Transformer as the architecture. Here, $L_p$ is the persistence length, which describes the bending stiffness of the chain. $L_p=0$ means no bending energy, i.e., the adjacent three beads are free to bend. The X-marks indicate that the model is unable to predict such length without truncation of the polymer; the em-dashes represent the absence of data.}

\begin{ruledtabular}
\begin{tabular}{c c c c c c c c | c}
Model & Feature & $L_p$ & $N = 100$ & $N=200$ & $N=300$ & $N=512$ & $N=1000$ & \pmb{$\forall N\leq 1000$} \\
\hline

Bi-LSTM (Vandas et al. 2020) & Bond vector & $10a$ & 99.59\% & \xmark & \xmark  & \xmark & \xmark& \xmark\\
Bi-LSTM (Braghetto et al. 2023) & Bond vector & \textbf{0}& \textemdash & \textemdash & \textemdash &  80\% \footnote{Polymers are partitioned into $N'=128$.} & \xmark & \xmark  \\
Bi-LSTM (Sleiman et al. 2024) & Segment writhe & $10a$ &99.8\% & \textbf{99.6\%} & \xmark& \xmark& \xmark & \xmark\\
\textbf{Transformer} (This paper) & Coordinates + Bond vector & \textbf{0} & \textbf{99.9999\% } & 99.08\%  & \textbf{99.9986\%} & \textbf{99.23\%} & \textbf{97.91\%} & \textbf{99.39\%}\footnote{On test set.} \\
\end{tabular}
\end{ruledtabular}
\end{table*}

We use the Transformer-based NN  (Fig.\ \ref{fig:Schematics}b) to recognize knot types (Fig.\ \ref{fig:Schematics}a) and achieve high accuracy (Fig.\ \ref{fig:bigmodel}). We produce the chain conformations with different knot types using Langevin dynamics simulations \cite{zhu2022computational,Vandans1} (see the method section). These simulations start with ring chain conformations with given knot types, and the knot types remain unchanged during simulations. We produce $60000$ chain conformations for each chain length ($N$=100, 300, 500, 800, 1000) and each knot type ($0_1$, $3_1$, $4_1$, $5_1$, $5_2$, $6_1$, $6_2$, $6_3$). So, the total number is 2400000. Out of 60000 conformations, 45000 are used for training, 5000 for validation, and 10000 for testing. In the NN, the input feature includes the coordinates of chain beads and the bond vectors (Fig.\ \ref{fig:Schematics}b). The training typically takes 20 epochs. 

Note that all the models in previous studies \cite{Vandans1, Braghetto1,Sleiman1} can not handle variable chain lengths, where the knotted polymers' lengths are around $N=200$ or shorter (truncation is necessary for longer lengths). This is usually accompanied by the problem of weak generalization ability to unseen lengths. We tackle this limitation by using the Transformer architecture and providing more lengths in the training. Note that the chain lengths in the training dataset, $N$=100, 300, 500, 800, 1000, are chosen completely arbitrarily, meaning that one might choose different lengths to reach a similar or even better performance. Moreover, the maximum chain length $N=1000$ is solely limited by the device's memory. 

Table\,\ref{tab:table1} shows that our model achieves excellent accuracy on the test dataset for all lengths seen during the training, i.e., $N$=100, 300, 500, 800, 1000. Especially, for $N$=100, the accuracy is 99.9999\%. The average accuracy reaches 99.39\% (Table\,\ref{tab:table1}).

To take a step further, we test our model's generalization ability. We apply it to classify polymer knots of length $N=80,\,150,\,437,\,550,\,821,\,952$. These chain lengths are picked arbitrarily to ensure we obtain the model's true performance. Fig.\ \ref{fig:bigmodel} shows the confusion matrices. Our model performs well for these arbitrarily chosen chain lengths, showcasing a robust generalization ability toward variable chain lengths.

Then, we compare our model's performance with previous studies \cite{Vandans1, Braghetto1,Sleiman1} (Table\,\ref{tab:table1}). Previous studies usually adopted LSTM architecture. For polymer knots of length $N=100$, we surpass the previous models reaching nearly 100\% accuracy. However, our model does not perform as strongly when classifying $N=200$. Multiple factors lead to this underperformance. First is the random fluctuation inherent in flexible polymers. For polymers with large bending stiffness in previous studies, the segments are more consistent with their neighbors hence simplifying knot recognition and facilitating the learning process. Secondly, the under-performance is mainly due to inaccuracy in classifying $5_1$ knots ($\approx$ 95.2\%), while the average accuracy of classifying other knots reaches 99.7\%. This might be due to a low variance in the data we collect for the $5_1$ knots of length $N=200$ and can be resolved by training more high-quality data. For polymer knots over $N=200$, Bi-LSTM models lose the ability to capture the topology, and hence a modification of the data is needed. In the work by Braghetto et al.\cite{Braghetto1}, they truncated the polymer ring by using a representative bead among the neighboring beads while deleting the rest. The resulting accuracy they obtained for $N=512$ tops merely 80\%, while ours reaches 99.23\%.

\subsection{Comparison of machine learning and mathematical methods}

\begin{figure}
\centering
\includegraphics[width=\linewidth]{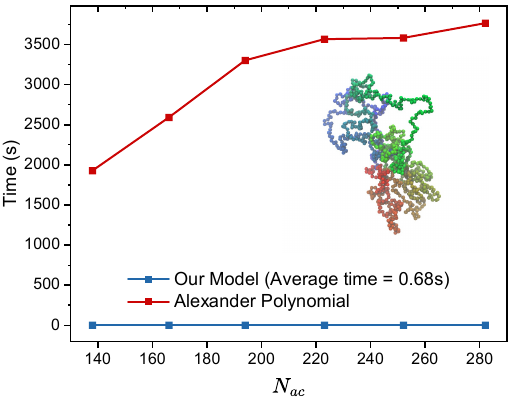}
\caption{\label{fig:comp_time} Comparison of total computational time of identifying $500$ polymer knots with high apparent crossing numbers $N_{ac}$ using traditional Alexander Polynomial and our NN model. Neural Network beats traditional Alexander Polynomial in speed. Inset shows a complex trefoil knot with $N_{ac} =258$ despite having a small minimum crossing number $N_c = 3$.}
\end{figure}

We compare the computational speeds of recognizing knot types using machine learning and mathematical methods. In the recent research of polymer knots, DNA knots, and protein knots, the most popular mathematical method to recognize knot types is the calculation of the Alexander polynomials for given conformations. We estimate the computational speeds based on a dataset of 500 polymer knots. For our machine learning model, the computational speed is 1.36 millisecond per polymer-knot conformation, while for the Alexander polynomials, the speed is  6.24 second per conformation (Fig.\,\ref{fig:comp_time}). So our speed is about 4500 times faster than the Alexander polynomials. Note that our computational speed is insensitive to the polymer length and the number of apparent crossings, while the calculation of Alexander polynomials strongly depends on the number of apparent crossings. The most time-consuming step in calculating the Alexander polynomials is computing the determinant of a matrix with $N_{ac} \times N_{ac}$, where $N_{ac}$ is the number of apparent crossings of a polymer conformation projected on a plane (see inset of Fig.\,\ref{fig:comp_time}). $N_{ac}$ is different from the number of crossings in the knot type, $N_{c}$, such as 3 for $3_1$ and 4 for $4_1$. Usually, $N_{ac}$ strongly depends on the polymer length and the compactness of polymer conformation. We can expect that the speed advantage of our method with respect to the Alexander polynomials becomes more obvious for longer polymers and denser packing of polymer conformations. In mathematics, there are other methods to recognizing knot types, such as HOMFLY polynomials. However, these methods are much slower than the Alexander polynomials. For example, the compuational cost for HOMFLY polynomials exponentially depends on the number of apparent crossings.   

\begin{figure*}
\centering
\includegraphics[width=\textwidth]{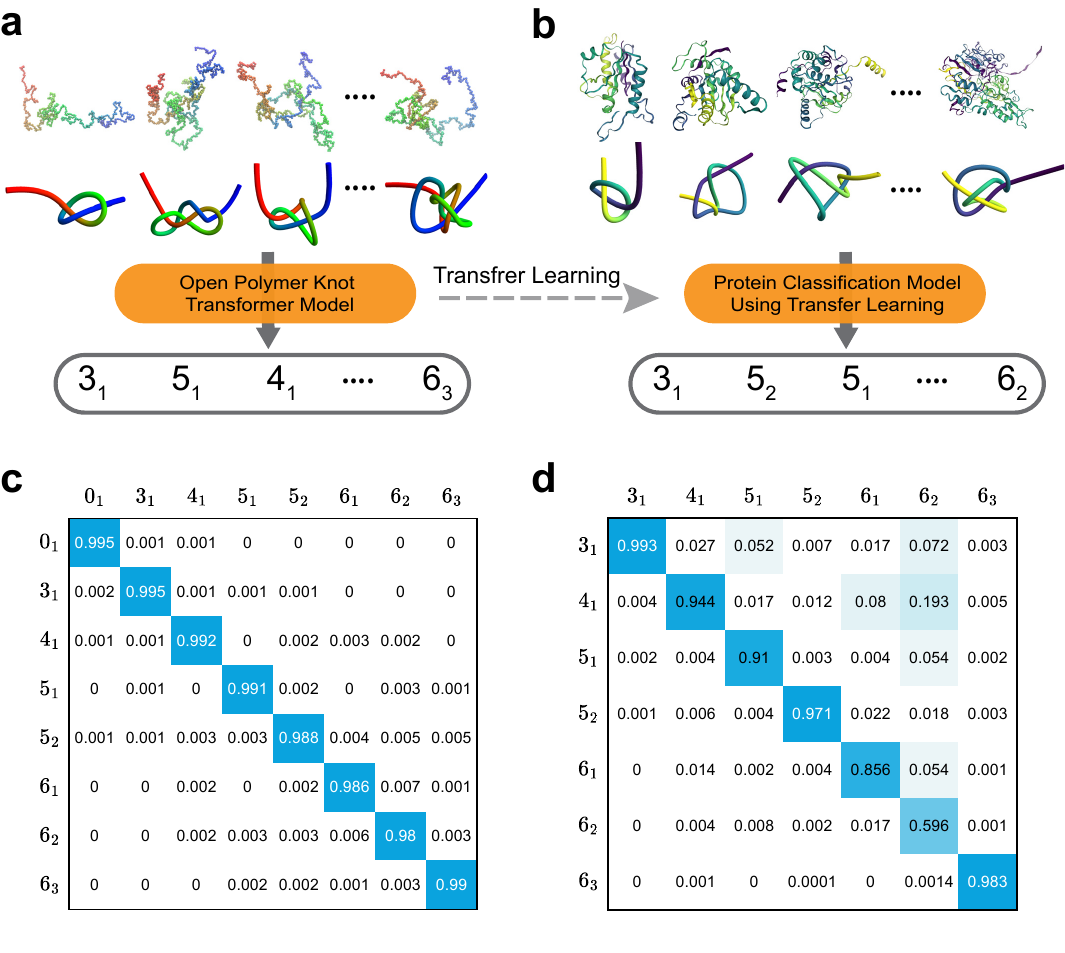}
\caption{\label{fig:protein_conf} Open polymer knots and protein knots classification. (a) Open polymer knots of various lengths. From left to right are representative polymers of length $N=574, \, 726, \, 828, \, 570$. We train a model to classify long linear polymer knots of different lengths up to $N=957$. (b) Realistic protein knots of various lengths. Protein Data Bank (PDB) codes, from left to right, are 1uam, 1xd3, Q54T48, and A0A133ZA81. We use transfer learning to extend our model from simple polymer knots to realistic protein knots classification. (c) Confusion matrices for linear knots. The average accuracy is nearly 99\%.  (d) Confusion matrices for protein knots. The classification result is obtained using transfer learning on the dataset from AlphaKnot. The average accuracy is approximately 82.4\%. For both matrices, the x coordinate represents the actual ground truth class, while the y coordinate represents the predicted.}
\end{figure*}
\subsection{Recognizing protein knots}
Then, we extend our model from simple polymer knots to realistic protein knots (Fig.\ \ref{fig:protein_conf}). This extension requires one modification: adapting the model from closed chain conformations (Fig.\ \ref{fig:Schematics}a) to open chain conformations (Fig.\ \ref{fig:protein_conf}a), because knotted proteins are usually open chains (without closure at the ends). Mathematically, knot type can be only defined for closed chains. For open chains, knot type is usually defined after closing the two chain ends by a loop \cite{tubiana2011probing}. The closing loop has been optimized to minimize the interference of the closing loop and chain conformation such that the determined knot type is usually consistent with human intuition.
To train an NN to recognize protein knots, we use the transfer learning: (i) first prepare a large dataset of open coarse-grained chain conformations to train an NN; (ii) then refine the NN by the dataset of knotted protein structures. It is because the protein knot dataset, AlphaKnot\cite{rubach2024alphaknot}, contains small numbers of complex knots. AlphaKnot contains 139,950 protein knots. The most abundant type is trefoil knots (90,846 instances), and the least is $6_2$ knots (166 instances).
In this first step, the dataset contains $10^6$ samples of eight knot types, with lengths ranging from 80 to 950. The model achieves an accuracy close to 99.9\% for open chain conformations (Fig.\ \ref{fig:protein_conf}a). In the second step, i.e., transfer learning, we use 20\% of AlphaKnot dataset as the training set. With a learning rate of $10^{-5}$ and training for only one epoch, we achieve an accuracy of 97.9\% (Fig.\ \ref{fig:protein_conf}b). 
For coarse-grained knotted chains, all knot types are identified with excellent accuracy, which means our model correctly captures the knot topology of open chains (Fig.\ \ref{fig:protein_conf}c). When extending our model to proteins using transfer learning, the classification still reaches great accuracy for $3_1, \,4_1 ,\, 5_1,\,5_2, \,6_1,$ and $6_3$ as shown in Fig.\ \ref{fig:protein_conf}d. For $6_2$, we observe a degenerate result. This is due to insufficient training data for $6_2$ protein knots.
\begin{figure*}
\centering
\includegraphics[width=\textwidth]{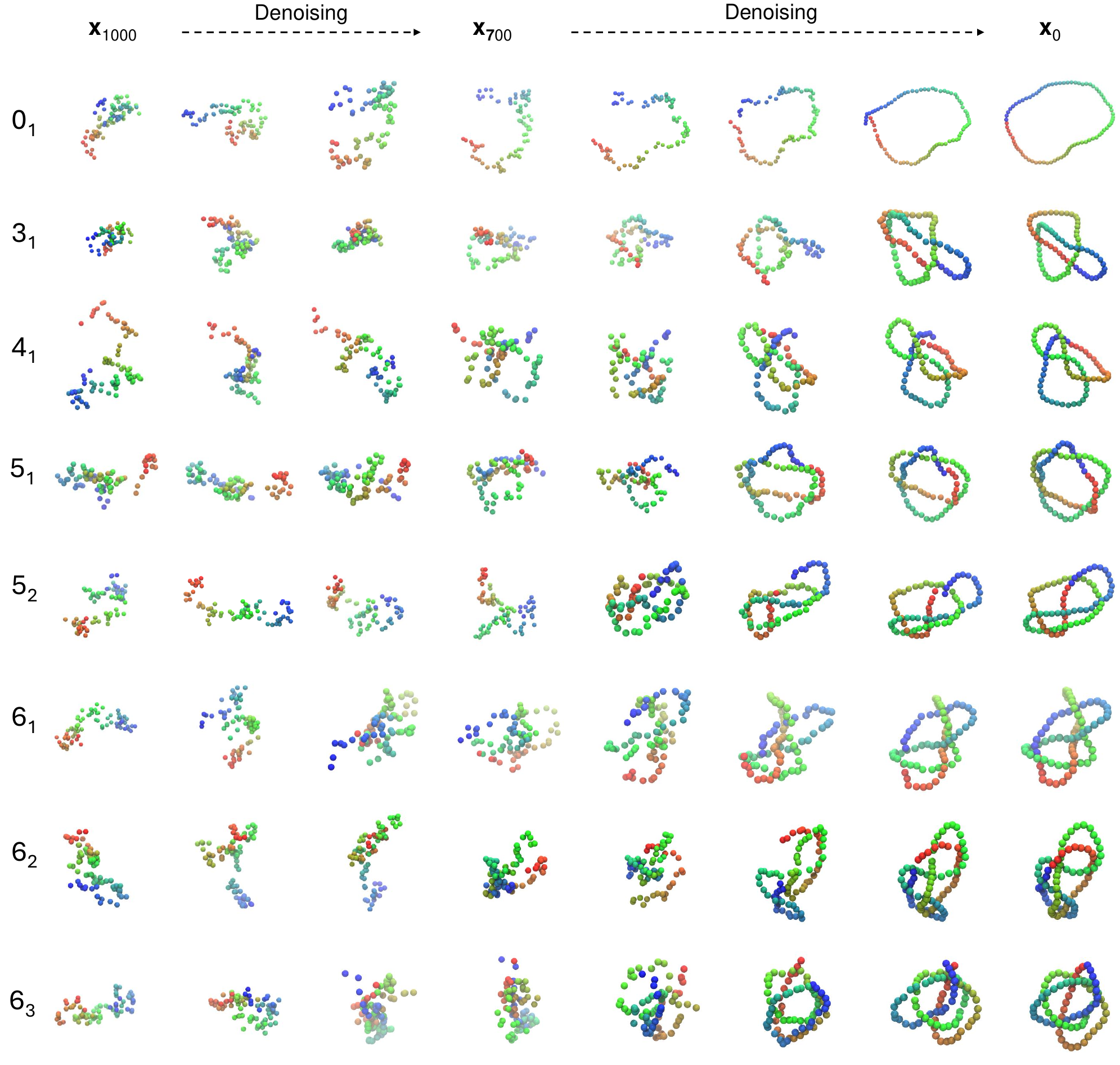}
\caption{\label{fig:diff_process}Diffusion process from normal distributed point clouds at $\mathbf{x}_{1000}$ to a structured knotted semi-flexible polymer at $\mathbf{x}_{0}$ for each knot type studied. The point clouds are converted from bond vector space to spatial coordinate space.}
\end{figure*}

\subsection{Generating knot structures by diffusion model}
\begin{figure*}
\centering
\includegraphics[width=\textwidth]{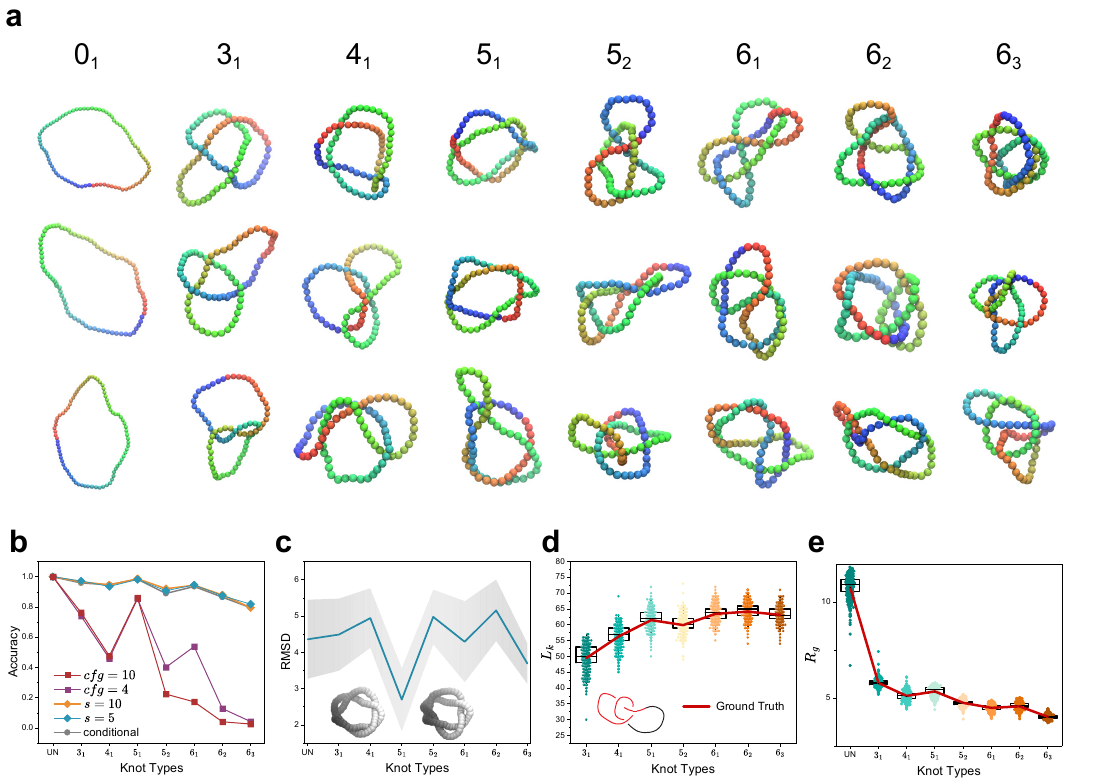}
\caption{\label{fig:poly_gene}Generation of knotted semi-flexible polymers through Diffusion model. (a) Examples of the generated knot conformations of different knot types. (b) Accuracy for different methods. $s$ and $cfg$ represent the guidance scale for CG and CFG respectively. (c) Root Mean Square Deviation (RMSD) between generated configurations and training dataset. The inset represents two knots with minimum RMSD. The knot on the left is the generated configuration. (d-e) Order parameters for different knot types generated using CG and $s=10$. The order parameters are compared to the real polymers as the ground truth. (d) The knot size for different knot types, which is defined as the contour length of knot core, illustrated by the red curve of the cartoon. (e) The radius of gyration for different knot types including the unknot.}
\end{figure*}

Next, we generate chain conformations for given knot types using the diffusion model (Fig.\ \ref{fig:Schematics}c). We adopt the training procedure in the Denoising Diffusion Probabilistic Model (DDPM), but replace the usual U-Net architecture with the Transformer due to its excellence in the knot classification task presented above. To train the model generating chain conformations with the desired knot types, we incorporate the knot invariant as the \textit{condition}. To encode our topological information as the condition, we pass it through multiple cross-attention layers to integrate with noised input embeddings (Fig.\ \ref{fig:Schematics}c). We train our model on a dataset of semi-flexible ring polymer knots of length $N=80$ with bond vectors as the feature (see supplementary for flexible polymer knots and ablation study on min-max normalized XYZ).

We schedule the noise for $1000$ steps, whereby the generation process corresponds to denoising from $T=1000$ to $T=0$ as shown in Fig.\ \ref{fig:diff_process}. At $T=1000$, i.e., the first step of the generation process,  the conformations are obtained from randomly distributed 3D point clouds of bond vectors. These bond vectors can be converted to spatial coordinates through cumulative vector sum. 
For the generation process, we adopt three methods: conditional generation, Classifier-Guidance (CG) \cite{dhariwal2021diffusion}, and Classifier Free Guidance (CFG) \cite{ho2022classifier}. For the conditional generation, we input topological invariant (knot type) to the model to guide the generation of knotted structures. For CG, we train a separate classifier that learns noisy knots' topology under different scheduling time steps. The log probability gradient of the classifier output is added to the learned mean of noise to sample (Fig. \ref{fig:Schematics}c). Lastly, for CFG, the implicit classifier is integrated into the training process by randomly selecting 10\% data trained without condition.

We evaluate the quality of generated structures in four terms (Fig.\ \ref{fig:poly_gene}): (i) the accuracy in the term of achieving the desired knot type; (ii) the generated structures are distinct from the ones in the training dataset, i.e., the novelty of the generated structures; (iii) the generated structures are significantly different from each other, i.e., the diversity of the generated structures; (iv)  whether the generated knotted structures follow realistic physical distributions, i.e. the generated structures are not only mathematically (topologically) correct but also physically correct.

We find that both conditional (CG guidance $s=0$) and CG generation achieve higher accuracy than CFG (Fig.\ \ref{fig:poly_gene}b). The accuracy for more complex knots is lower. This is reasonable because a more complex knot has smaller conformational space, i.e., smaller conformational entropy in statistical physics. It is easy to see in molecular simulations that random chain conformations have a lower probability to adopt a more complex knot \cite{dai2014metastable,micheletti2011polymers}. We find the incorrect knot structure generated by our model has certain pattern that bears mathematical reasoning. For example, the most incorrect structures generated for $6_3$ are $3_1$, because the $6_3$ knot can be readily converted to a $3_1$ knot if a crossing is moved from under-passing to over-passing (see supplementary).

To evaluate the novelty of the generated knotted structures, we calculate the RMSD (root mean square deviation) between the generated structures and the training dataset (Fig.\ \ref{fig:poly_gene}c). For each generated structure, we search the closest structure in the training dataset and record the RMSD value for this generated structure. The RMSD value and the thickness of the shadow region for each knot type in Fig.\ \ref{fig:poly_gene}c are the average and standard deviation for the generated structures in a specific knot type. The RMSD values are typically around 4 (in the unit of bead diameter), which indicates significant difference between generated and training structures. The two closest structures in the generated and training structures are presented in the inset of Fig.\ \ref{fig:poly_gene}c.

To show the diversity of the generated structures, we present three examples for each knot type (Fig.\ \ref{fig:poly_gene}a). In addition, the thickness of the shadow region in Fig. \ref{fig:poly_gene}c also indicates the diversity of the generated structures.

To evaluate whether the generated structures follow correct physical distributions, we first analyze the contour length of the knot core, $L_k$, (Fig.\ \ref{fig:poly_gene}d). The knot core refers to the minimum portion of the chain that maintains the knot type (see the inset of Fig.\ \ref{fig:poly_gene}d). We can determine the knot core by cutting the ring chain to an open chain and then cutting beads from both chain ends until the change in the knot type. Here, $L_k$ is simply the number of beads in the knot core. A small $L_k$ means the knot is localized on the chain, while a large $L_k$ means the knot spreads over the entire chain. Calculation and analysis of $L_k$ have been extensively performed in our previous studies \cite{Dai1, lu2024knotting, zhu2021revisiting}, because $L_k$ is an important property for a knotted structure. Fig.\ \ref{fig:poly_gene}d shows the generated structures agree with the training dataset in terms of $L_k$, where the knotted structures in the training dataset are produced by Langevin dynamics simulations and follow the correct physical distribution.
 

Another order parameter associated with polymer conformation is the radius of gyration $R_g$, which is a quantity that represents the extension of the polymer in 3D and is highly anti-correlated with the complexity of knot types. Fig.\ \ref{fig:poly_gene}e shows the generated structures agree with the training dataset in terms of $R_g$, which means the generated structures have the physically correct conformational compactness. As the minimum crossing number increases, the knots become more complicated and hence absorb more segments into the knotted region in the polymer rings, leading to a decrease in the total size (Fig.\ \ref{fig:poly_gene}a and e). $R_g$ can also be theoretically determined for unknots but not easily determined for knotted polymers. Here, we compare the generated configurations to our dataset ground truth for different knot types. Both generated polymers and molecular simulations show the same decreasing trend in $R_g$ and the values agree (Fig.\ \ref{fig:poly_gene}e).

\section{Discussion}
\subsection{Discussion about protein knot generation}
Accurate and robust generation is crucial to both \textit{de novo} protein design\cite{Trippe1,Ingraham1,Notin1} and recent AI-MS models\cite{Noe1,Arts1,Zhang2,Hsu1,Zhang1}. This entails that the statistical order parameters, structures, and topology should agree with the real physical world. 

Chroma is a recent \textit{de novo} design model that generates proteins in a programmable manner, which is achieved by using certain conditioner energy and running Markov Chain Monte Carlo sampling to reach the desired protein structure \cite{Ingraham1}. However, designing such conditioner energy functions for knots requires intricate thinking. A possible detour is generating knotted proteins from existing ones, by fixing the knot core while letting the model generate the rest. Here we select three knotted proteins (1UAM, 1NS5, and 1XD3) from the Protein Data Bank (PDB) with known knot core positions. After fixing residue positions corresponding to the knotted core, generated knotted proteins have different structures (see supplementary).

We propose that in that, in the future, a Large Protein Model (LPM) similar to the Large Language Models (LLMs) and Large Vision Models (LVMs) can use our technique to generate specific topologies such as knots. This can be done by training a smaller model, such as Contrastive Language-Image Pretraining (CLIP) \cite{CLIP} for instance, which can be used to guide the generation of the desired topology without retraining the big model.

\subsection{Why the Transformer excels in knotting tasks}

Our results demonstrate the success of applying the Transformer in recognizing and generating knotted structures. Here, we briefly discuss the reason for this success. The knot type strongly depends on the seqeunce of the bead positions of a chain. For comparison, the radius of gyration depends on the assembly of the bead positions, but not the sequence. Therefore, using NN architecture that is strong at sequence processing to handle knotting tasks is necessary. It is known that LSTM and Transformer are capable of processing sequence information. Hence, previous studies using LSTM and this work using Transformer perform well in handling knotting tasks. One likely reason why the Transformer performs better than LSTM is that the Transformer can handle long-term memory better than LSTM. In LSTM, the long-term memory decays with the distance between two tokens along the sequence, or the contour difference between two beads along a chain. In the Transformer, the attention mechanism can relate long-distance tokens by matching the query and key. In the case of knotting, a crossing often comes from two beads that are far away from each other along the chain but close to each other in 3D space. A slight shift in a bead position can switch the under-over status of a crossing and then change the knot type. This situation is amplified in long polymers, for which LSTM becomes overshadowed by Transformer as we have shown in the main text. Hence, precise and efficient processing of the long-distance correlation is critical for knotting tasks, which is the strength of the Transformer.     

\subsection{Discussion about feature selection}

In our previous work, we used only the bond vector as the feature for knot topology. In this work, using spatial coordinates alone is sufficient, while including the bond vector improves the result in the Transformer model (see supplementary). Interestingly, when training using the Transformer (Fig\,\ref{fig:Schematics}b), the bond vector alone fails to capture the topology (see supplementary). However, by adding a Multilayer Perceptron (MLP) before the Transformer block, we find that the result is drastically improved to a similar accuracy as in our main text. This is reflected in RNN-based LSTM model as it learns the topology sequentially through encoding each token using an MLP layer. Hence, we argue that the NN learns knot topology by first converting the bond vector to spatial coordinate, i.e. $ \mathbf{x_i} = \sum^i_0 \mathbf{b_j}$, or some implicit representation which is easier for the NN to capture the topology hidden in the spatial and sequential information of the beads. A similar situation happens with feature embeddings such as the segment-to-segment writhe (StS) and the integral of segment-to-segment writhe (StA) utilized in a previous study\cite{Sleiman1}. StA fails to represent the topology in flexible polymers due to large thermal fluctuations among the segments because the integral erases some important sequential information in the polymer. On the contrary, StS correctly represents the topology in flexible polymers as it conserves the sequential order by calculating the writhe at each monomer site.

\subsection{Discussion about future improvement}

Our result shows the attention mechanism effectively learns polymer knots and encodes topology to guide the generative process. We use mostly the same values for hyper-parameters in our models as those in the original Transformer and Diffusion model papers, as opposed to a carefully selected set of hyperparameters in our first paper\cite{Vandans1}. Careful tuning of these parameters might result in an even better overall accuracy for our model in the classification task. Moreover, including shorter lengths in the training stage of our model can improve generalizing accuracy for smaller polymers. It is worth noting that previous literature has done data manipulation like truncation to incorporate long polymers. However, this may undermine the original topological information and hence lead to a poor result. 

For our generative models, generation accuracy is not as stunning as that of our classification model. For future design of a model, one may need to consider a more sophisticated way of incorporating prior knowledge of knot invariants into training. For example, as mentioned previously in the main text, knot types are extremely sensitive to over-under passing of segments. Hence, such prior knowledge of under and over-passing can be used to regularize generation. To obtain useful prior knowledge, one may use the KMT algorithm to first reduce knots into simpler configurations and project them into a 2-D plane to avoid the model learning unimportant high-level features. Integrating this prior knowledge in the training stage of the latent Diffusion model could largely help the NN encode the knot topology in the latent space.  


\section{Conclusion}

To conclude, our work shows that the Transformer is a successful improvement of the current Bi-LSTM model in learning long polymer knots by careful selection of input features. In the task of classification, with the help of the highly efficient computing power of the Transformer architecture, we resolve the problem of length generalization by training multiple lengths at the same time. We use various lengths at the training stage to train a model that classifies 8 different types of knots with excellent accuracy, and it is completely feasible to include more complex knots. For even better accuracy, more configurations and lengths can be included into the training dataset. We also train a separate model on open polymer knots and reach similar accuracy in classifying open-chain knots. Since protein knots are also mostly open chains, we extend this model to protein knots using transfer learning and obtained great accuracy. 

For the second task, we, for the first time, investigate and develop a model to robustly generate accurate semi-flexible polymer knots. We use knot invariant as the condition for our generation and compared three different methods of generation. We find that both conditional and Classifier-Guided diffusion generate knots most accurately, in contrast to Classifier-Free Guidance. We also examine two order parameters of the generated knots and find they both agree with the physical ground truth.

Our Tranformer-based and diffusion-based models achieve good performance in recognizing and generating knotted structures. The impact of our results is multi-fold.

First, our models solve three critical issues in applying the machine learning methods to recognize knot types, which are heavily needed in the research of polymer knots, DNA knots, and protein knots. The three issues are: variable chain lengths, a wide range of knot types, and high accuracy. Machine learning models in previous studies are limited to a fixed chain length. It means that different models are needed for polymer knots with different chain lengths, which is inconvenient and impractical. Here, our model performs well for all chain lengths not larger than 1000. In the research of polymer knots, DNA knots, and protein knots, the chain lengths are mostly less than 1000. In addition, the maximum chain length can be increased by training models on machines with more memory.

Second, we for the first time demonstrate that machine learning methods can generate knotted structures with high accuracy. Generation of knotted structures has great practical values, because knotted structures have more special properties, such as special catalysis and extraordinary structural stability. In this work, we apply the knot generation for coarse-grained polymer chains. It is expected that the model can be integrated in protein design models for generating knotted protein structures using auxiliary models such as CLIP \cite{CLIP} .

Third, our results show that the Transformer is a very suitable machine learning architecture to handle knots or broadly, topology. It is well known that different machine learning architecture are suitable for different tasks, such as CNN for image tasks and RNN for time-series tasks. Our results that the Transformer outperforms the LSTM-based RNN in the task regarding knots. The outstanding performance of Transformer in this task is reasonable considering the knotting problem (a string of beads positions) is implicit in the tokens which can be efficiently solved through parallel computing.

Last but not least, knotting problem is not fully solved in math. One issue is the many-to-one mapping, i.e. for conformations with different knot types, the corresponding Alexander polynomials and other polynomials may be same. Analyzing how good machine learning models recognize knot types may provide insights into the principle behind knots. One recent study has demonstrated that machine learning helps mathematicians to reveal a new connection between the algebraic and geometric structure of knots \cite{davies2021advancing}.

\section{Methods}
\subsection{Preparation of the dataset of polymer knots using molecular simulations}

We generated our dataset of polymer conformations using Langevin dynamics simulations to model flexible polymer rings, consisting of connected monomer beads with diameter $\sigma$, each undergoes thermal motion \cite{zhu2022computational}. The beads interact through hard-core collisions, enforced by the Weeks-Chandler-Andersen (WCA) potential. To model the bond energy, we use the Kremer-Grest model\cite{kremergrest}, where the finitely extensible nonlinear elastic (FENE) potential is applied to lock the topology of the knotted polymer ring. The net potential $\mathrm{V_{tot}} = \mathrm{V_{FENE}} + \mathrm{V_{\mathrm{excl}}} $ hence has two parts, where the first part is from the FENE potential $\mathrm{V_{FENE}}$ $$\mathrm{V_{FENE}} = -30\epsilon\frac{\mathrm{r}^2_0}{\sigma^2}   \ln \left ( 1-\Bigl(\frac{\mathrm{r}}{\mathrm{r}_0}\Bigr)^2\right)$$ with $\mathrm{r}_0 = 1.5\sigma$ and $\epsilon = 1$.  The second part is the Weeks-Chandler-Andersen (WCA) exclusion potential $\mathrm{V_{\mathrm{excl}}}$ among neighboring beads is
\begin{equation}
  \mathrm{V_{\mathrm{excl}}} =
    \begin{cases}
      4\epsilon \left[(\frac{\sigma}{\mathrm{r}})^{12} - (\frac{\sigma}{\mathrm{r}})^6) \right] + \epsilon &  \mathrm{r} < 2^{\frac{1}{6}} \sigma\\
      0 & \text{otherwise.}
    \end{cases}       
\end{equation}  
The resulting equation of motion for each monomer bead is hence
\begin{equation}
 m\frac{d{\mathbf{v}_i}(t)}{dt}=-\gamma \mathrm{v_i}(t) - \sum_j{ \nabla  \mathrm{V_{tot}}(\bf r_i - \bf r_j) } + \mathrm{\mathbf{f}}_\mathrm{i}^\mathrm{r}(\mathrm{t}),
\end{equation}
where $m$ is the mass, $\gamma$ is the drag coefficient, and $\mathrm{\mathbf{f}}_\mathrm{{i}}^\mathrm{{r}}$ is a Gaussian random force term that averages zero. The simulation is performed by solving this equation of motion with the velocity-Verlet algorithm in the NVT ensemble using LAMMPS. We set $\gamma=\sigma=m=\epsilon=T=1$ and run simulations in free space for $10^8$ steps with step size $\Delta t = 10^{-2}\tau$, where $\tau =\sqrt{\sigma^2m/\epsilon}=1$. 

In this study, we considered knots with up to six minimum crossings. In this work, we train a model using polymer lengths of $N=100,\,300,\,500,\,800,\, 1000$. Fig.\ \ref{fig:Schematics}a presents some representative configurations and corresponding knot diagrams. Note the knot structures are challenging to discern visually due to significant segmental fluctuations, resulting from the absence of bending energy.

\subsection{Machine learning for recognizing knots}


We use a six-dimensional feature vector composed of XYZ coordinates and bond vectors to represent the topology of the polymer knots. We deploy the Transformer encoder architecture to learn the embedded feature of long polymer knots. Since the attention mechanism inherently ignores the sequential information, a positional embedding is needed to mark the sequential order of the beads. Here, we choose the sinusoidal embedding as in the original paper by Vaswani et al\cite{Vaswani1}. The sum of the feature embedding and the positional embedding is sent to the Transformer encoder. 

Inside the Transformer encoder is multiple multi-headed attention layers, where, in each layer, each bead's summed embedding is attended with the other beads' and itself, a technique called the self-attention mechanism. The attention is done by first passing the summed feature embedding through a multi-layer perception (MLP), which augments (or shrinks) the dimension of the embedding so that the resulting tensor can be separated into three tensors: Query (Q), Key (K), and Value (V). Then the dot product of the Q and K tensor is passed through another MLP and then a soft-max function, and then the V tensor is dotted with the resulting tensor. The dot product enables parallel computation and long-range feature learning. In the final stage of our model, the encoded features are then passed down through a linear layer, and the class (CLS) token is used for the knot classification. Fig.\ \ref{fig:Schematics}B summarizes our model. Note that we used hyper-parameters for Transformer layers same as Vaswani et al\cite{Vaswani1}, indicating that our embedding features and model are powerful in describing and learning the flexible polymer knots in free space, and tuning these hyper-parameters may further improve the accuracy. 

\subsection{Machine learning for generating knots}
To generate polymer knots, we use a modified Diffusion model. First, a noise scheduling that adds Gaussian noise at each step $T$ is applied to the feature vectors of the polymer knots. The resulting noisy coordinates are passed to a denoiser composed of several self-attention and cross-attention layers to learn the noise, which will be subtracted when generating the knots. The self-attention layers encode the input feature vectors, and the cross-attention layers integrate topological invariants into each bond as the condition. To ensure our model generates in the correct sequential order, we use sinusoidal positional embedding to embed the position of beads. We further encoded the noise scheduling time with Gaussian Random Fourier Features to capture high-frequency information\cite{tancik2020fourier} and adopt the AdaNorm \cite{xu2019understanding} to integrate the time features into the attention layers. 

For the final generation, we adopt three methods: conditional generation, Classifier-Guidance (CG) \cite{dhariwal2021diffusion}, and Classifier Free Guidance (CFG) \cite{ho2022classifier}. For the conditional generation, we input topological invariant to the model to generate knots. For CG, we train a separate classifier that learns noisy knots' topology under different scheduling time steps. The log probability gradient of the classifier output is added to the learned mean of noise to sample (see Fig.\ \ref{fig:Schematics}c and supplementary for details).

We implement our models using Pytorch. Our code and model parameters can be accessed on \href{https://github.com/kizzhang/KnotTransformer}{GitHub}.


\begin{acknowledgments}
This research is financially supported by National Natural Science Foundation of China (project no. 22273080), Research Grants Council of Hong Kong (project no. 11313322, 11307224), CityU Institute of Digital Medicine, and Guangdong Basic and Applied Basic Research Fund (project no. 2022A1515010484).
\end{acknowledgments}

\bibliographystyle{unsrt}
\bibliography{Trans_knot}

\begin{thebibliography}{10}

\bibitem{Krasnow1}
M.~A. Krasnow, A.~Stasiak, S.~J. Spengler, F.~Dean, T.~Koller, and N.~R. Cozzarelli.
\newblock Determination of the absolute handedness of knots and catenanes of dna.
\newblock {\em Nature}, 304(5926):559--60, 1983.

\bibitem{Rybenkov1}
V.~V. Rybenkov, N.~R. Cozzarelli, and A.~V. Vologodskii.
\newblock Probability of dna knotting and the effective diameter of the dna double helix.
\newblock {\em Proc Natl Acad Sci U S A}, 90(11):5307--11, 1993.

\bibitem{Rybenkov2}
Valentin~V Rybenkov, Christian Ullsperger, Alexander~V Vologodskii, and Nicholas~R Cozzarelli.
\newblock Simplification of dna topology below equilibrium values by type ii topoisomerases.
\newblock {\em Science}, 277(5326):690--693, 1997.

\bibitem{Arsuaga1}
Javier Arsuaga, Mariel Vázquez, Sonia Trigueros, and Joaquim Roca.
\newblock Knotting probability of dna molecules confined in restricted volumes: Dna knotting in phage capsids.
\newblock {\em Proc Natl Acad Sci U S A}, 99(8):5373--5377, 2002.

\bibitem{Plesa1}
Calin Plesa, Daniel Verschueren, Sergii Pud, Jaco van~der Torre, Justus~W Ruitenberg, Menno~J Witteveen, Magnus~P Jonsson, Alexander~Y Grosberg, Yitzhak Rabin, and Cees Dekker.
\newblock Direct observation of dna knots using a solid-state nanopore.
\newblock {\em Nat. Nanotech.}, 11:1093–1097, 2016.

\bibitem{Taylor1}
William~R Taylor.
\newblock A deeply knotted protein structure and how it might fold.
\newblock {\em Nature}, 406:916--919, 2000.

\bibitem{Mallam1}
Anna~L Mallam and Sophie~E Jackson.
\newblock Knot formation in newly translated proteins is spontaneous and accelerated by chaperonins.
\newblock {\em Nat. Chem. Biol.}, 8(2):147, 2012.

\bibitem{San1}
Álvaro San~Martin, Piere Rodriguez-Aliaga, Jose~Alejandro Molina, Andreas Martin, Carlos Bustamante, and Mauricio Baez.
\newblock Knots can impair protein degradation by atp\-dependent proteases.
\newblock {\em Proc Natl Acad Sci USA}, 114:9864, 2017.

\bibitem{Rubach1}
Pawel Rubach, Maciej Sikora, Aleksandra~I Jarmolinska, Agata~P Perlinska, and Joanna~I Sulkowska.
\newblock Alphaknot 2.0: a web server for the visualization of proteins’ knotting and a database of knotted alphafold-predicted models.
\newblock {\em Nucleic Acids Res.}, page gkae443, 2024.

\bibitem{Sogo1}
José~M Sogo, Andrzej Stasiak, Marıa~Luisa Martınez-Robles, Dora~B Krimer, Pablo Hernández, and Jorge~B Schvartzman.
\newblock Formation of knots in partially replicated dna molecules.
\newblock {\em J Mol. Bio.}, 286(3):637--643, 1999.

\bibitem{Liu1}
Zhirong Liu, Richard~W Deibler, Hue~Sun Chan, and Lynn Zechiedrich.
\newblock The why and how of dna unlinking.
\newblock {\em Nucleic acids research}, 37(3):661--671, 2009.

\bibitem{Marenduzzo1}
Davide Marenduzzo, Cristian Micheletti, and Enzo Orlandini.
\newblock Topological friction strongly affects viral dna ejection.
\newblock {\em Proc Natl Acad Sci U S A}, 110(50):20081--20086, 2013.

\bibitem{Vologodskii1}
Alexander Vologodskii.
\newblock Disentangling dna molecules.
\newblock {\em Phys. Life Rev.}, 18:118--134, 2016.

\bibitem{Schvartzman1}
Jorge~B Schvartzman, Pablo Hernández, Dora~B Krimer, Julien Dorier, and Andrzej Stasiak.
\newblock Closing the dna replication cycle: from simple circular molecules to supercoiled and knotted dna catenanes.
\newblock {\em Nucleic acids res.}, 47(14):7182--7198, 2019.

\bibitem{Berger1}
James~M Berger, Steven~J Gamblin, Stephen~C Harrison, and James~C Wang.
\newblock Structure and mechanism of dna topoisomerase ii.
\newblock {\em Nature}, 379(6562):225--232, 1996.

\bibitem{Nitiss1}
John~L Nitiss.
\newblock Targeting dna topoisomerase ii in cancer chemotherapy.
\newblock {\em Nat. Rev. Cancer}, 9(5):338--350, 2009.

\bibitem{Jamroz1}
Michal Jamroz, Wanda Niemyska, Eric~J Rawdon, Andrzej Stasiak, Kenneth~C Millett, Piotr Sułkowski, and Joanna~I Sulkowska.
\newblock Knotprot: a database of proteins with knots and slipknots.
\newblock {\em Nucleic acids res.}, 43(D1):D306--D314, 2014.

\bibitem{Sulkowska1}
Joanna~I Sułkowska, Piotr Sułkowski, P~Szymczak, and Marek Cieplak.
\newblock Stabilizing effect of knots on proteins.
\newblock {\em Proc Natl Acad Sci}, 105:19714, 2008.

\bibitem{zhu2022computational}
Haoqi Zhu, Fujia Tian, Liang Sun, Yongjian Zhu, Qiyuan Qiu, and Liang Dai.
\newblock Computational design of extraordinarily stable peptide structures through side-chain-locked knots.
\newblock {\em Journal of Physical Chemistry Letters}, 13(33):7741--7748, 2022.

\bibitem{Christian1}
Thomas Christian, Reiko Sakaguchi, Agata~P Perlinska, Georges Lahoud, Takuhiro Ito, Erika~A Taylor, Shigeyuki Yokoyama, Joanna~I Sulkowska, and Ya-Ming Hou.
\newblock Methyl transfer by substrate signaling from a knotted protein fold.
\newblock {\em Nature structural \& molecular biology}, 23(10):941--948, 2016.

\bibitem{Marcos1}
Vanesa Marcos, Alexander~J Stephens, Javier Jaramillo-Garcia, Alina~L Nussbaumer, Steffen~L Woltering, Alberto Valero, Jean-François Lemonnier, Iñigo~J Vitorica-Yrezabal, and David~A Leigh.
\newblock Allosteric initiation and regulation of catalysis with a molecular knot.
\newblock {\em Science}, 352(6293):1555--1559, 2016.

\bibitem{Tang1}
Jing Tang, Ning Du, and Patrick~S Doyle.
\newblock Compression and self-entanglement of single dna molecules under uniform electric field.
\newblock {\em Proc Natl Acad Sci U S A}, 108:16153--16158, 2011.

\bibitem{Renner1}
C~Benjamin Renner and Patrick~S Doyle.
\newblock Stretching self-entangled dna molecules in elongational fields.
\newblock {\em Soft Matter}, 11(16):3105--3114, 2015.

\bibitem{Zhang3}
Min Zhang, Robert Nixon, Fredrik Schaufelberger, Lucian Pirvu, Guillaume De~Bo, and David~A Leigh.
\newblock Mechanical scission of a knotted polymer.
\newblock {\em Nat. Chem.}, pages 1--7, 2024.

\bibitem{Coluzza1}
Ivan Coluzza, Peter~DJ van Oostrum, Barbara Capone, Erik Reimhult, and Christoph Dellago.
\newblock Sequence controlled self-knotting colloidal patchy polymers.
\newblock {\em Phys. Rev. Lett.}, 110(7):075501, 2013.

\bibitem{Sharma1}
Rajesh~Kumar Sharma, Ishita Agrawal, Liang Dai, Patrick Doyle, and Slaven Garaj.
\newblock Dna knot malleability in single-digit nanopores.
\newblock {\em Nano Lett.}, 21(9):3772--3779, 2021.

\bibitem{Sharma2}
Rajesh~K Sharma, Ishita Agrawal, Liang Dai, Patrick~S Doyle, and Slaven Garaj.
\newblock Complex dna knots detected with a nanopore sensor.
\newblock {\em Nat. Commun.}, 10:4473, 2019.

\bibitem{Suma1}
Antonio Suma and Cristian Micheletti.
\newblock Pore translocation of knotted dna rings.
\newblock {\em Proc Natl Acad Sci U S A}, 114:E2991, 2017.

\bibitem{Reifenberger1}
Jeffrey~G Reifenberger, Kevin~D Dorfman, and Han Cao.
\newblock Topological events in single molecules of e. coli dna confined in nanochannels.
\newblock {\em Analyst}, 140(14):4887--4894, 2015.

\bibitem{Mao1}
Runfang Mao and Kevin~D Dorfman.
\newblock Dynamics of double-knotted dna molecules under nanochannel confinement.
\newblock {\em Macromolecules}, 2024.

\bibitem{Ma1}
Zixue Ma and Kevin~D Dorfman.
\newblock Diffusion of knotted dna molecules in nanochannels in the extended de gennes regime.
\newblock {\em Macromolecules}, 54(9):4211--4218, 2021.

\bibitem{Alexander1}
James~W Alexander.
\newblock Topological invariants of knots and links.
\newblock {\em Transactions of the American Mathematical Society}, 30(2):275--306, 1928.

\bibitem{Jones1}
Vaughan~FR Jones.
\newblock {\em Hecke algebra representations of braid groups and link polynomials}, pages 20--73.
\newblock World Scientific, 1987.

\bibitem{Freyd1}
Peter Freyd, David Yetter, Jim Hoste, WB~Raymond Lickorish, Kenneth Millett, and Adrian Ocneanu.
\newblock A new polynomial invariant of knots and links.
\newblock {\em Bull. Am. Math. Soc.}, 12:239, 1985.

\bibitem{Ozsváth1}
Peter Ozsváth and Zoltán Szabó.
\newblock Holomorphic disks and knot invariants.
\newblock {\em Advances in Mathematics}, 186(1):58--116, 2004.

\bibitem{Khovanov1}
Mikhail Khovanov.
\newblock A categorification of the jones polynomial.
\newblock 2000.

\bibitem{Hochreiter1}
Sepp Hochreiter and Jürgen Schmidhuber.
\newblock Long short-term memory.
\newblock {\em Neural computation}, 9(8):1735--1780, 1997.

\bibitem{Vandans1}
Olafs Vandans, Kaiyuan Yang, Zhongtao Wu, and Liang Dai.
\newblock Identifying knot types of polymer conformations by machine learning.
\newblock {\em Physical Review E}, 101(2):022502, 2020.

\bibitem{Braghetto1}
Anna Braghetto, Sumanta Kundu, Marco Baiesi, and Enzo Orlandini.
\newblock Machine learning understands knotted polymers.
\newblock {\em Macromolecules}, 56(7):2899--2909, 2023.

\bibitem{Sikora1}
Maciej Sikora, Eva Klimentova, Dawid Uchal, Denisa Sramkova, Agata~P Perlinska, Mai~Lan Nguyen, Marta Korpacz, Roksana Malinowska, Szymon Nowakowski, and Pawel Rubach.
\newblock Knot or not? identifying unknotted proteins in knotted families with sequence‐based machine learning model.
\newblock {\em Protein Science}, 33(7):e4998, 2024.

\bibitem{Sleiman1}
Joseph~Lahoud Sleiman, Filippo Conforto, Yair Augusto~Gutierrez Fosado, and Davide Michieletto.
\newblock Geometric learning of knot topology.
\newblock {\em Soft Matter}, 20(1):71--78, 2024.

\bibitem{Wang1}
Yu~Wang, Luwei Lu, Zhenhua Wang, Guoqiang Zhou, and Yuyuan Lu.
\newblock Integrating graph convolutional network to bilstm for enhanced polymer knot identification.
\newblock {\em Macromolecules}, 57(16):7980--7989, 2024.

\bibitem{Tubiana1}
Luca Tubiana, Angelo Rosa, Filippo Fragiacomo, and Cristian Micheletti.
\newblock Spontaneous knotting and unknotting of flexible linear polymers: Equilibrium and kinetic aspects.
\newblock {\em Macromolecules}, 46(9):3669--3678, 2013.

\bibitem{Orlandini1}
Enzo Orlandini.
\newblock Statics and dynamics of dna knotting.
\newblock {\em J. Phys. A}, 51(5):053001, 2017.

\bibitem{Dai1}
Liang Dai.
\newblock Tube model for polymer knots with excluded volume interactions and its applications.
\newblock {\em Macromolecules}, 54(20):9299--9306, 2021.

\bibitem{Bengio1}
Y.~Bengio, P.~Simard, and P.~Frasconi.
\newblock Learning long-term dependencies with gradient descent is difficult.
\newblock {\em IEEE Transactions on Neural Networks}, 5(2):157--166, 1994.

\bibitem{Vaswani1}
Ashish Vaswani, Noam Shazeer, Niki Parmar, Jakob Uszkoreit, Llion Jones, Aidan~N Gomez, Łukasz Kaiser, and Illia Polosukhin.
\newblock Attention is all you need.
\newblock {\em Advances in neural information processing systems}, 30, 2017.

\bibitem{Dosovitskiy1}
Alexey Dosovitskiy, Lucas Beyer, Alexander Kolesnikov, Dirk Weissenborn, Xiaohua Zhai, Thomas Unterthiner, Mostafa Dehghani, Matthias Minderer, Georg Heigold, and Sylvain Gelly.
\newblock An image is worth 16x16 words: Transformers for image recognition at scale.
\newblock {\em arXiv preprint arXiv:2010.11929}, 2020.

\bibitem{Radford1}
Alec Radford, Karthik Narasimhan, Tim Salimans, and Ilya Sutskever.
\newblock Improving language understanding by generative pre-training.
\newblock 2018.

\bibitem{Brown1}
Tom Brown, Benjamin Mann, Nick Ryder, Melanie Subbiah, Jared~D Kaplan, Prafulla Dhariwal, Arvind Neelakantan, Pranav Shyam, Girish Sastry, and Amanda Askell.
\newblock Language models are few-shot learners.
\newblock {\em Advances in neural information processing systems}, 33:1877--1901, 2020.

\bibitem{Trippe1}
Brian~L Trippe, Jason Yim, Doug Tischer, David Baker, Tamara Broderick, Regina Barzilay, and Tommi Jaakkola.
\newblock Diffusion probabilistic modeling of protein backbones in 3d for the motif-scaffolding problem.
\newblock {\em arXiv preprint arXiv:2206.04119}, 2022.

\bibitem{Ingraham1}
J.~B. Ingraham, M.~Baranov, Z.~Costello, K.~W. Barber, W.~Wang, A.~Ismail, V.~Frappier, D.~M. Lord, C.~Ng-Thow-Hing, E.~R. Van~Vlack, S.~Tie, V.~Xue, S.~C. Cowles, A.~Leung, J.~V. Rodrigues, C.~L. Morales-Perez, A.~M. Ayoub, R.~Green, K.~Puentes, F.~Oplinger, N.~V. Panwar, F.~Obermeyer, A.~R. Root, A.~L. Beam, F.~J. Poelwijk, and G.~Grigoryan.
\newblock Illuminating protein space with a programmable generative model.
\newblock {\em Nature}, 623(7989):1070--1078, 2023.

\bibitem{Notin1}
Pascal Notin, Nathan Rollins, Yarin Gal, Chris Sander, and Debora Marks.
\newblock Machine learning for functional protein design.
\newblock {\em Nature Biotechnology}, 42(2):216--228, 2024.

\bibitem{Zhang2}
Linfeng Zhang, Jiequn Han, Han Wang, Roberto Car, and Weinan E.
\newblock Deepcg: Constructing coarse-grained models via deep neural networks.
\newblock {\em The Journal of Chemical Physics}, 149(3), 2018.

\bibitem{Noe1}
Frank Noé, Simon Olsson, Jonas Köhler, and Hao Wu.
\newblock Boltzmann generators: Sampling equilibrium states of many-body systems with deep learning.
\newblock {\em Science}, 365(6457):eaaw1147, 2019.

\bibitem{Arts1}
Marloes Arts, Victor Garcia~Satorras, Chin-Wei Huang, Daniel Zugner, Marco Federici, Cecilia Clementi, Frank Noé, Robert Pinsler, and Rianne van~den Berg.
\newblock Two for one: Diffusion models and force fields for coarse-grained molecular dynamics.
\newblock {\em Journal of Chemical Theory and Computation}, 19(18):6151--6159, 2023.

\bibitem{Hsu1}
Tim Hsu, Babak Sadigh, Vasily Bulatov, and Fei Zhou.
\newblock Score dynamics: Scaling molecular dynamics with picoseconds time steps via conditional diffusion model.
\newblock {\em Journal of Chemical Theory and Computation}, 20(6):2335--2348, 2024.

\bibitem{Virnau1}
P.~Virnau, L.~A. Mirny, and M.~Kardar.
\newblock Intricate knots in proteins: Function and evolution.
\newblock {\em PLoS Comput Biol}, 2(9):e122, 2006.

\bibitem{Kolesov1}
Grigory Kolesov, Peter Virnau, Mehran Kardar, and Leonid~A Mirny.
\newblock Protein knot server: detection of knots in protein structures.
\newblock {\em Nucleic acids research}, 35(suppl\_2):W425--W428, 2007.

\bibitem{Lou1}
Shih-Chi Lou, Svava Wetzel, Hongyu Zhang, Elizabeth~W Crone, Yun-Tzai Lee, Sophie~E Jackson, and Shang-Te~Danny Hsu.
\newblock The knotted protein uch-l1 exhibits partially unfolded forms under native conditions that share common structural features with its kinetic folding intermediates.
\newblock {\em Journal of molecular biology}, 428(11):2507--2520, 2016.

\bibitem{tubiana2011probing}
Luca Tubiana, Enzo Orlandini, and Cristian Micheletti.
\newblock Probing the entanglement and locating knots in ring polymers: a comparative study of different arc closure schemes.
\newblock {\em Progress of Theoretical Physics Supplement}, 191:192--204, 2011.

\bibitem{rubach2024alphaknot}
Pawel Rubach, Maciej Sikora, Aleksandra~I Jarmolinska, Agata~P Perlinska, and Joanna~I Sulkowska.
\newblock Alphaknot 2.0: a web server for the visualization of proteins’ knotting and a database of knotted alphafold-predicted models.
\newblock {\em Nucleic Acids Research}, page gkae443, 2024.

\bibitem{dhariwal2021diffusion}
Prafulla Dhariwal and Alexander Nichol.
\newblock Diffusion models beat gans on image synthesis.
\newblock {\em Advances in neural information processing systems}, 34:8780--8794, 2021.

\bibitem{ho2022classifier}
Jonathan Ho and Tim Salimans.
\newblock Classifier-free diffusion guidance.
\newblock {\em arXiv preprint arXiv:2207.12598}, 2022.

\bibitem{dai2014metastable}
Liang Dai, C~Benjamin Renner, and Patrick~S Doyle.
\newblock Metastable tight knots in semiflexible chains.
\newblock {\em Macromolecules}, 47(17):6135--6140, 2014.

\bibitem{micheletti2011polymers}
Cristian Micheletti, Davide Marenduzzo, and Enzo Orlandini.
\newblock Polymers with spatial or topological constraints: Theoretical and computational results.
\newblock {\em Physics Reports}, 504(1):1--73, 2011.

\bibitem{lu2024knotting}
Luwei Lu, Qiyuan Qiu, Yuyuan Lu, Lijia An, and Liang Dai.
\newblock Knotting in flexible-semiflexible block copolymers.
\newblock {\em Macromolecules}, 2024.

\bibitem{zhu2021revisiting}
Haoqi Zhu, Fujia Tian, Liang Sun, Simin Wang, and Liang Dai.
\newblock Revisiting the non-monotonic dependence of polymer knotting probability on the bending stiffness.
\newblock {\em Macromolecules}, 54(4):1623--1630, 2021.

\bibitem{Zhang1}
J.~Zhang, D.~Chen, Y.~Xia, Y.~P. Huang, X.~Lin, X.~Han, N.~Ni, Z.~Wang, F.~Yu, L.~Yang, Y.~I. Yang, and Y.~Q. Gao.
\newblock Artificial intelligence enhanced molecular simulations.
\newblock {\em J Chem Theory Comput}, 19(14):4338--4350, 2023.

\bibitem{CLIP}
Alec Radford, Jong~Wook Kim, Chris Hallacy, Aditya Ramesh, Gabriel Goh, Sandhini Agarwal, Girish Sastry, Amanda Askell, Pamela Mishkin, Jack Clark, Gretchen Krueger, and Ilya Sutskever.
\newblock Learning transferable visual models from natural language supervision.
\newblock In Marina Meila and Tong Zhang, editors, {\em Proceedings of the 38th International Conference on Machine Learning}, volume 139 of {\em Proceedings of Machine Learning Research}, pages 8748--8763. PMLR, 18--24 Jul 2021.

\bibitem{davies2021advancing}
Alex Davies, Petar Veli{\v{c}}kovi{\'c}, Lars Buesing, Sam Blackwell, Daniel Zheng, Nenad Toma{\v{s}}ev, Richard Tanburn, Peter Battaglia, Charles Blundell, Andr{\'a}s Juh{\'a}sz, et~al.
\newblock Advancing mathematics by guiding human intuition with ai.
\newblock {\em Nature}, 600(7887):70--74, 2021.

\bibitem{kremergrest}
Kurt Kremer and Gary~S Grest.
\newblock Dynamics of entangled linear polymer melts: A molecular-dynamics simulation.
\newblock {\em The Journal of Chemical Physics}, 92(8):5057--5086, 1990.

\bibitem{tancik2020fourier}
Matthew Tancik, Pratul Srinivasan, Ben Mildenhall, Sara Fridovich-Keil, Nithin Raghavan, Utkarsh Singhal, Ravi Ramamoorthi, Jonathan Barron, and Ren Ng.
\newblock Fourier features let networks learn high frequency functions in low dimensional domains.
\newblock {\em Advances in neural information processing systems}, 33:7537--7547, 2020.

\bibitem{xu2019understanding}
Jingjing Xu, Xu~Sun, Zhiyuan Zhang, Guangxiang Zhao, and Junyang Lin.
\newblock Understanding and improving layer normalization.
\newblock {\em Advances in neural information processing systems}, 32, 2019.

\end{thebibliography}

\end{document}